\documentclass{aastex61}
\usepackage{graphicx}

\usepackage[modulo, pagewise]{lineno}
\usepackage{ulem}
\usepackage{color,soul}

\begin{document}
\title{Detection of A Water Tracer in Interstellar Comet 2I/Borisov}
\correspondingauthor{Adam McKay}
\email{adam.mckay@nasa.gov}
\author{Adam J. McKay}
\affil{NASA Goddard Space Flight Center/American University}
\author{Anita L. Cochran} 
\affil{McDonald Observatory, University of Texas at Austin}
\author{Neil Dello Russo}
\affil{Johns Hopkins Applied Physics Laboratory}
\author{Michael A. DiSanti}
\affil{NASA Goddard Space Flight Center}


\begin{abstract}
We present high spectral resolution optical spectra obtained with the ARCES instrument at Apache Point Observatory showing detection of the [\ion{O}{1}]6300~\AA~line in interstellar comet 2I/Borisov.  We employ the observed flux in this line to derive an H$_2$O production rate of (6.3$\pm$1.5)$\times$10$^{26}$ mol s$^{-1}$.  Comparing to previously reported observations of CN, this implies a CN/H$_2$O ratio of $\sim$0.3-0.6\%. The lower end of this range is consistent with the average value in comets, while the upper end is higher than the average value for Solar System comets, but still within the range of observed values.  C$_2$/H$_2$O is depleted, with a value likely less than 0.1\%.  The dust-to-gas ratio is consistent with the normal value for Solar System comets.  Using a simple sublimation model we estimate an H$_2$O active area of 1.7 km$^2$, which for current estimates for the size of Borisov suggests active fractions between 1-150\%, consistent with values measured in Solar System comets.  More detailed characterization of 2I/Borisov, including compositional information and properties of the nucleus, is needed to fully interpret the observed H$_2$O production rate.
\end{abstract}

\section{Introduction}
\indent Comets have a primitive volatile composition that is thought to reflect the conditions present in their formation region in the protosolar disk.  This makes studies of cometary volatiles powerful for understanding the physical and chemical processes occurring during planet formation.  However, observations to date only sample comets with a Solar System origin.  While the possibility exists that some comets with anomolous compositions such as 96P/Macholz~\citep{LanglandShulaSmith2007, Schleicher2008} and C/2016 R2 (PanSTARRS)~\citep{Biver2018,McKay2019} could have interstellar origins, the dynamics of these comets do not provide conclusive proof that they originate from a star system other than our own.\\
\indent The discovery of interstellar comet 2I/Borisov provides an opportunity to sample the volatile composition of a comet that is unambiguously from outside our own Solar System, providing constraints on the physics and chemistry of other protostellar disks.  So far the only volatile that has been conclusively detected in Borisov is CN, with upper limits reported for C$_2$~\citep{Fitzsimmons2019, Kareta2019}, C$_3$, and OH~\citep{Opitom2019b}, and constraints on C$_2$ showing it is likely depleted compared to Solar System comets.  As H$_2$O is the dominant volatile in most Solar System comets, measuring the H$_2$O production in Borisov is key for interpretation of all other observations of this comet, including other volatiles.\\  
\indent The [\ion{O}{1}]6300~\AA~line can be used as a proxy for the H$_2$O production rate in comets~\citep[e.g][]{Morgenthaler2007,Fink2009,McKay2018}.  This line is a forbidden transition whose upper state is most efficiently populated when \ion{O}{1} is released into the coma in the $^1$D state after photodissociation of a parent molecule.  In cometary comae for the $^1$D state, this parent molecule is usually H$_2$O.  We present observations of the [\ion{O}{1}]6300~\AA~line in Borisov and employ these observations to provide a measure of the H$_2$O production rate. 

\section{Observations and Analysis}
\indent We obtained spectral observations of Borisov on UT October 11, 2019 using the ARCES instrument mounted on the 3.5-meter Astrophysical Research Consortium (ARC) Telescope at Apache Point Observatory in Sunspot, NM.  ARCES is a cross-dispersed, high spectral resolution spectrograph with continuous spectral coverage from 3500-10,000~\AA~and a resolving power of R $\equiv$ $\frac{\lambda}{\delta\lambda}$=31,500.  This high resolving power is necessary for observations of [\ion{O}{1}]6300~\AA~emission in order to separate the cometary line from the corresponding telluric feature (see Fig.~\ref{OI}).  More information about the ARCES instrument can be found elsewhere~\citep{Wang2003}.\\
\indent Details of the observations can be found in Table~\ref{observations}.  We obtained two spectra with 1800 second exposure times at airmasses of 2.14 and 1.74.  The slit was oriented lengthwise along the parallactic angle. We observed the fast rotating A star HD 80613 to serve as a telluric standard and HR 3454 for flux calibration of the spectra.  Both were observed at airmass $\sim$2.0, similar to the airmass of the Borisov observations.  Hyades 64 was also observed as a solar standard for removal of solar absorption features, but due to the extremely weak nature of the continuum in the spectra of Borisov and the absence of any detectable absorption features we decided not to apply our solar standard to the cometary observations.  We obtained observations of a quartz lamp for flat fielding and a ThAr lamp for wavelength calibration.  Spectra were reduced using an IRAF script that performs bias subtraction, flat fielding, cosmic ray removal, spectral extraction, and wavelength calibration.  More details of our reduction procedures for cometary ARCES data can be found in~\cite{McKay2012, McKay2014}.  After reduction and calibration, we co-added the two cometary spectra to increase signal-to-noise.\\

\begin{table}[h!]
\begin{center}
\caption{\textbf{Observation Log}
\label{observations}
}
\begin{tabular}{cccccccc}
\hline
UT Date & R$_h$ (AU) & $\dot{R_h}$ (km s{$^{-1}$}) &$\Delta$ (AU) & $\dot{\Delta}$ (km s{$^{-1}$}) & Solar Standard & Tell. Standard & Flux Standard\\
\hline
10/11/19 & 2.38 & -20.5 & 2.81 & -34.2 & Hyades 64 & HD 80613 & HR 3454\\
\end{tabular}
\end{center}
\end{table}

\indent ARCES has a small slit (3.2$\times$1.6\arcsec).  Therefore we accounted for slit losses from our flux standard using aperture photometry on our slitviewer images and the methodology of~\cite{McKay2014}.  For these observations slit losses add a systematic uncertainty in the derived flux of 12\%.\\
\indent We retrieved the observed flux by fitting a Gaussian to the line profile and calculating the corresponding flux.  This flux is then used in a Haser model that includes modifications that emulate the vectorial formalism~\citep{Festou1981} and also accounts for collisional quenching of the [\ion{O}{1}]6300~\AA~line emission.  More details of this model can be found in~\cite{McKay2012, McKay2014}.\\

\section{Results and Discussion}
\begin{figure}[h!]
\includegraphics[width=\textwidth]{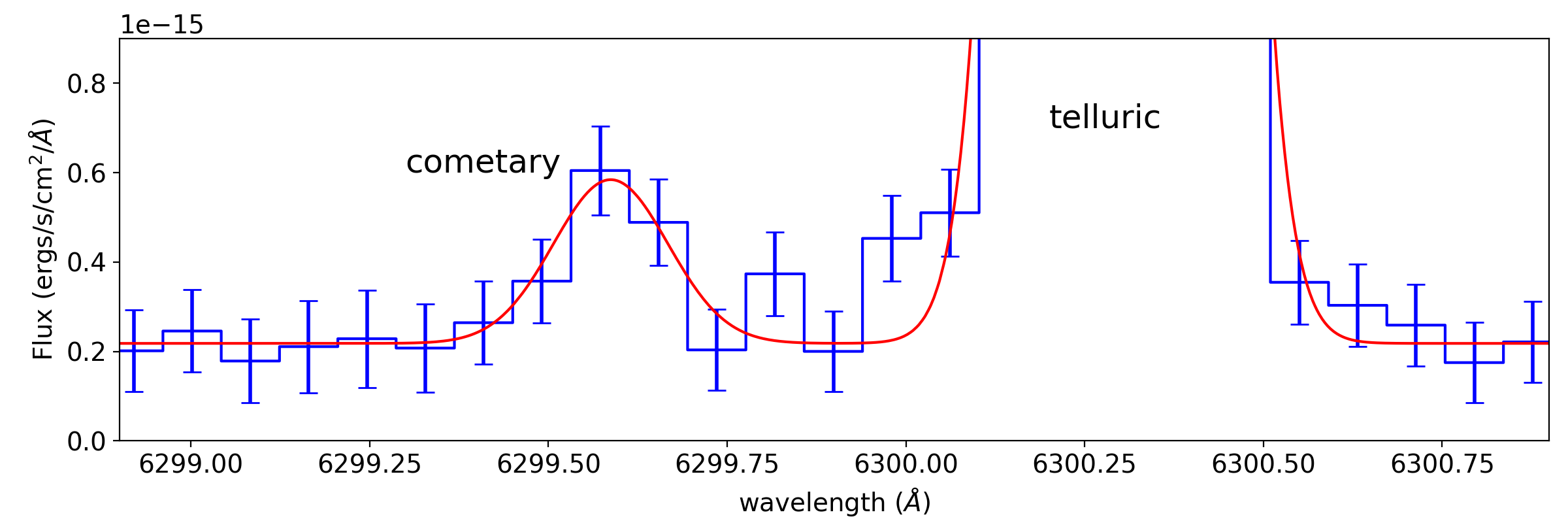}
\caption{Spectral region showing the [\ion{O}{1}]6300~\AA~line in Borisov for our co-added spectrum (1 hour total on-source integration time).  The telluric feature extends off the top of the plot, with the cometary line being the weaker feature blueward of the telluric line.  The red line shows Gaussian fits to both the cometary and telluric features.  These spectra do not have the continuum removed, so the continuum level was also a fitting parameter.  Error bars represent $\pm$1$\sigma$ uncertainties.
}
\label{OI}
\end{figure}

\indent Figure~\ref{OI} shows our detection of the [\ion{O}{1}]6300~\AA~line in Borisov, with a fit to both the cometary and telluric lines overplotted in red.   As the continuum baseline was not removed during reduction, we included the continuum level as an additional parameter during the Gaussian fitting process.  The cometary line is detected at the 5.0$\sigma$ level.  Assuming that H$_2$O is the dominant source of the  [\ion{O}{1}]6300~\AA~line emission, we derive an H$_2$O production rate of (6.3$\pm$1.5) $\times$ 10$^{26}$ mol s$^{-1}$.  The uncertainty is dominated by photon statistics in the spectra rather than by the uncertainty in flux calibration.  While our spectra cover other species of interest such as CN and C$_2$, no other emissions were detected and the upper limits derived from our observations would not provide more sensitive constraints than those already reported~\citep{Fitzsimmons2019,Kareta2019,Opitom2019b}.\\

\begin{figure}[h!]
\includegraphics[width=0.5\textwidth]{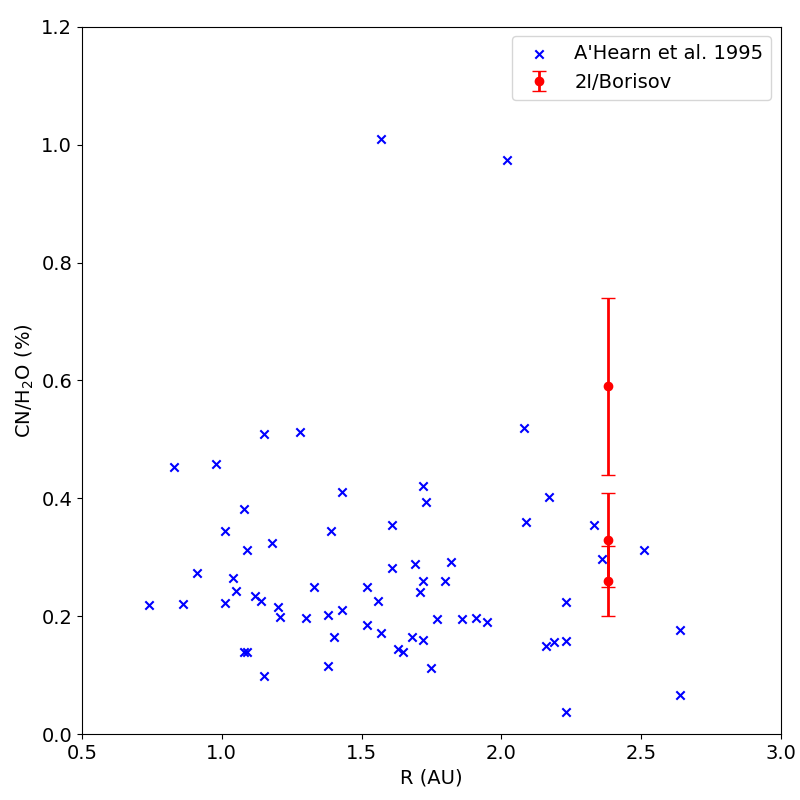}
\includegraphics[width=0.5\textwidth]{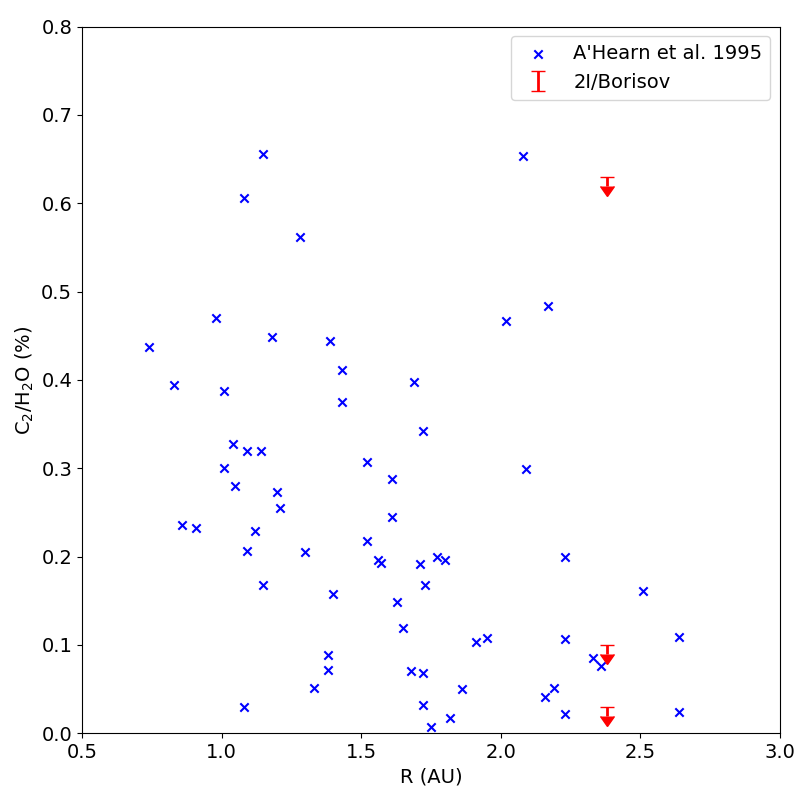}
\caption{Left: CN/H$_2$O ratio in comets observed by~\cite{AHearn1995} (blue x's) as a function of heliocentric distance, with results for Borisov overplotted in red.  As~\cite{AHearn1995} report CN/OH ratios, we have converted to CN/H$_2$O using the relation from~\cite{CochranSchleicher1993}.  The values based on the measurements by~\cite{Kareta2019} and~\cite{Opitom2019b} show values consistent with the mean value for Solar System comets and other comets observed at this heliocentic distance, while the ratio based on the~\cite{Fitzsimmons2019} measurement suggests a higher than average abundance, but still within the observed range for Solar System comets. Right: Same but for C$_2$/H$_2$O.  The same conversion from C$_2$/OH to C$_2$/H$_2$O as for CN was performed.  All values are upper limits, but the most sensitive values suggest Borisov is depleted in C$_2$.
}
\label{comparison}
\end{figure}

\indent \cite{Opitom2019b} report an upper limit on the OH production rate (another proxy for H$_2$O production) of 2.0 $\times$ 10$^{27}$ mol s$^{-1}$.  Our [\ion{O}{1}]6300~\AA~line detection is consistent with this upper limit.  While this manuscript was in review, Crovisier et al., in an IAU CBET, reported a tentative water production rate of (3.3$\pm$0.9) $\times$ 10$^{27}$ mol s$^{-1}$ based on observations of the 18-cm OH line with the Nan\c{c}ay radio telescope co-added from 15 hours of observations obtained over a period of three weeks.  This value is a factor of five higher than our value.  Possible reasons for this discrepancy are discussed at the end of this section.\\
\indent \cite{Fitzsimmons2019},~\cite{Kareta2019}, and~\cite{Opitom2019b} report detections of the CN molecule in the coma of Borisov.  Using their measured CN production rates for the most contemporaneous observations, our H$_2$O production rate implies a CN/H$_2$O ratio of 0.59 $\pm$ 0.15\%~\citep{Fitzsimmons2019}, 0.26 $\pm$ 0.06\%~\citep{Kareta2019}, or 0.33 $\pm$ 0.08\%~\citep{Opitom2019b}.  The CN mixing ratio derived from the~\cite{Fitzsimmons2019} CN production rate is higher than the mean value of Solar System comets measured to date~\citep{AHearn1995,Cochran2012}, but still within the range of values found for Solar System comets.  The CN values from~\cite{Opitom2019b} and~\cite{Kareta2019} give CN/H$_2$O ratios consistent both with each other and the mean value for Solar System comets.  As the measurements of \cite{Opitom2019b} and \cite{Kareta2019} are the most contemporaneous with our measurements (closest observation dates are UT October 13 and UT October 10, respectively), perhaps these observations provide the most relevant comparison in accounting for possible variability in outgassing.  We graphically compare the CN/H$_2$O ratio in Borisov to the sample of observed comets in~\cite{AHearn1995} in the left panel of Fig.~\ref{comparison}.\\
\indent For C$_2$, the upper limit from \cite{Fitzsimmons2019} combined with our result gives C$_2$/H$_2$O $<$ 0.63\%, consistent with the mean value for Solar System comets.  Using the upper limits from~\cite{Opitom2019b} and~\cite{Kareta2019} results in C$_2$/H$_2$O $<$ 0.1\% and $<$ 0.03\%, respectively, which is depleted compared to Solar System comets~\citep{AHearn1995,Cochran2012}.  We include a graphical comparison of the C$_2$/H$_2$O ratio in Borisov to the sample of observed comets in~\cite{AHearn1995} in the right panel of Fig.~\ref{comparison}.  More observations are needed as Borisov approaches perihelion in order to provide a more definitive measure of its C$_2$ abundance.\\

\indent ~\cite{Fitzsimmons2019} calculated an Af$\rho$ parameter, which serves as a proxy for dust production, of 143$\pm$10 cm from their observations.~\cite{Opitom2019b} determined a very similar number.  Using our measured H$_2$O production rate results in a log(Af$\rho$/$Q_{H_2O}$) value of -24.6, which is similar to Solar System comets observed at heliocentric distances similar to these observations ($\sim$ 2.5 AU)~\citep{AHearn1995}.~~\cite{JewittLuu2019} calculated from photometric observations and a dust model that the dust mass loss rate is approximately 2 kg s$^{-1}$, while~\cite{Fitzsimmons2019} derived a value of 1 kg s$^{-1}$.  Converting our H$_2$O production rate to mass results in an H$_2$O mass loss rate of $\sim$ 19 kg s$^{-1}$.  This is about a factor of three smaller than the rate calculated by~\cite{JewittLuu2019} and~\cite{Fitzsimmons2019} using the ~\cite{Fitzsimmons2019} CN production rate and assuming a typical CN/H$_2$O ratio for Solar System comets.  Our result that CN is enhanced in Borisov (at least when the~\cite{Fitzsimmons2019} CN production rate is employed) compared to the mean CN/H$_2$O ratio in Solar System comets explains this discrepancy.  This is also supported by the fact that if the CN production rate from~\cite{Opitom2019b} or~\cite{Kareta2019} is adopted (which compared with our H$_2$O production rate suggests a more typical CN/H$_2$O ratio), the inferred H$_2$O mass loss rate is $\sim$ 30 kg s$^{-1}$, in better agreement with our value based on direct detection of a water tracer.  Therefore, we conclude that the gas mass loss rate is about an order of magnitude larger than the dust mass loss rate, which would make Borisov incredibly gas-rich compared to the average Solar System comet, but similar to gas-rich endmembers like 2P/Encke~\citep{AHearn1995}.  However, it should be noted that dust mass loss rates are extremely model dependent, and~\cite{Fitzsimmons2019} noted that using 20$\mu$m grains instead of 1$\mu$m grains resulted in a dust mass loss rate of $\sim$30 kg s$^{-1}$, which would imply a dust to gas mass ratio closer to unity, consistent with the average value in Solar System comets.\\
\indent It is important to note that our observations are not simultaneous with those of~\cite{Fitzsimmons2019} and~\cite{JewittLuu2019}, so any variability in the outgassing behavior of Borisov would complicate interpretation of these mixing ratios.  In general, these previously reported observations were at larger heliocentric distance, where it would be expected that gas/dust production would be less.  Therefore, it is possible that the CN, C$_2$, and dust production concurrent with our observations may be larger than the numbers reported for earlier observations, meaning all mixing ratios may be somewhat higher.  However, the \cite{Opitom2019b} and \cite{Kareta2019} observations bracket ours and show a fairly constant CN production rate over this time period, so as mentioned earlier perhaps these observations provide the most relevant comparison to our observations.\\
\indent We use a simple sublimation model based on~\cite{CowanAHearn1979} in order to convert our measured H$_2$O production rate into an active area.  We assume properties typical of Solar System comets: low thermal inertia (justifying the slow rotator approximation, where every facet of the nucleus surface is in thermal equilibrium with the solar radiation incident upon it) and an albedo of 0.04.  We also assume a spherical nucleus for simplicity.  With these parameters we find an active area of 1.7 km$^2$.  Using the derived radius upper limit of 3.8 km from~\cite{JewittLuu2019} results in a lower limit on the active fraction of $\sim$1\%, consistent with Solar System comets~\citep{SosaFernandez2011,Lis2019}.  However, ~\cite{JewittLuu2019} argue that the nucleus is likely much smaller than this, perhaps only a few hundred meters in radius.~~\cite{Jewitt2019} using HST observations come to a similar conclusion, constraining the nucleus size to 200-500 meters in radius.  For a 300-meter body, our simple sublimation model implies an active fraction of 140\%, which would imply a hyperactive nucleus.  This phenomenon has been observed for Solar System comets~\citep{AHearn2011,Lis2019} and is often explained as resulting from an extended source of H$_2$O-rich ice grains being released into the coma.  \cite{Fitzsimmons2019} estimated based on their modeling efforts that the nucleus is 0.7-3.3 km in radius, resulting in active fractions of $\sim$1-25\%, similar to Solar System comets.  It must be noted however that our sublimation model uses a very simple treatment that is sensitive to thermal properties of the nucleus and albedo, which are not well constrained for Borisov.  A more detailed modeling approach is beyond the scope of this Letter, but would be very beneficial for understanding the properties and activity of Borisov.\\
\indent A possible complication in interpretation of the observed [\ion{O}{1}]6300~\AA~emission is that other volatiles such as CO, CO$_2$, and O$_2$ may contribute significantly to the emission if they are present at levels equal to or more abundant than H$_2$O.  A recent example of this is C/2016 R2 (PanSTARRS), for which CO$_2$ was 30 times more abundant and CO was 300 times more abundant than H$_2$O, respectively, and it was determined that CO and CO$_2$ were the dominant contributors to the observed [\ion{O}{1}]6300~\AA~line flux~\citep{McKay2019}.  A possible way to constrain the contribution of CO and CO$_2$ to the [\ion{O}{1}] population is through observations of the [\ion{O}{1}]5577~\AA~line~\citep{FestouFeldman1981, BhardwajRaghuram2012,McKay2012,Decock2013}.  While our spectra cover this feature, it was not detected and the upper limit on the flux ratio of the [\ion{O}{1}]5577~\AA~and [\ion{O}{1}]6300~\AA~lines of $\sim$ 1.0 is not sensitive enough to rule out CO and CO$_2$ as major contributors to the observed [\ion{O}{1}]6300~\AA~line flux.  However, modeling of the dust coma by~\cite{JewittLuu2019} suggests that activity began at a heliocentric distance of 4.5 AU, much closer to the Sun than would be expected if CO$_2$ or CO was a dominant driver of activity and more consistent with H$_2$O sublimation.  If confirmed, the tentative detection of OH using the Nan\c{c}ay radio telescope and the derived H$_2$O production rate (Crovisier et al. 2019) would provide additional evidence that the [\ion{O}{1}]6300~\AA~emission we observe does originate from H$_2$O photodissociation and therefore is an accurate tracer for H$_2$O.  Observations of CO$_2$ at IR wavelengths and CO at either IR or sub-mm wavelengths are important for ruling out these potential contributors to the observed [\ion{O}{1}]6300~\AA~line flux.\\
\indent As stated earlier, during review of this manuscript Croviser et al. announced in a CBET a tentative water production rate approximately five times larger than our reported value.  While the brief nature of the CBET precludes a detailed comparison, we discuss some possible reasons for this discrepancy.  At the high airmass of these observations and the small dimensions of the ARCES slit, differential refraction can result in wavelength dependent slit loss, which can skew flux measurements.  However, this is not expected for [\ion{O}{1}]6300~\AA~emission because this feature is close to the guiding wavelength ($\sim$ 5500~\AA).  We confirmed that this is indeed negligible for [\ion{O}{1}]6300~\AA~emission based on observations of comet C/2012 S1 (ISON) that were performed at similarly high airmass with ARCES, and found that the production rates derived from the ISON [\ion{O}{1}]6300~\AA~measurements were consistent with values determined using other methods~\citep{McKay2018}.  Therefore we do not consider this or other airmass dependent phenomena as the reason for the discrepancy.  At certain geocentric velocities the cometary [\ion{O}{1}]6300~\AA~emission sits on top of a strong telluric absorption, and at high airmass inaccurate removal of this feature can result in a decrease in the measured flux and therefore production rate.  This was observed for C/2012 S1 (ISON)~\citep{McKay2018}.  However, the geocentric velocity of 2I/Borisov during our observations was $\sim$ -35 km s$^{-1}$, while the effect on observed [\ion{O}{1}]6300~\AA~line fluxes in comet ISON was only observed at geocentric velocities of $\sim$ -50 km s$^{-1}$.  Therefore this is also not a likely candidate to explain the discrepancy.  It is also possible that the activity is highly variable, and we observed Borisov at a minimum in activity, while the Nan\c{c}ay observations, which were co-added over three weeks of observations, provide a long term average production rate.  However, no such variability is observed for CN, with the CN production rate being fairly constant over a several week period~\citep{Kareta2019,Opitom2019b}.\\
\indent The presence of an extended source of H$_2$O production, usually explained by the presence of water-rich icy grains in the coma, could account for the discrepancy.  The main evidence for an extended source of H$_2$O production in cometary comae from ground-based observations is a dependence of derived production rates on the projected area of sky (aperture size) over which the production rate is measured.  This phenomenon was observed for comet C/2009 P1 (Garradd)~\citep{Combi2013,Bodewits2014, McKay2015}, with production rates measured with larger aperture sizes giving larger values than smaller apertures.  The Nan\c{c}ay beam size is very large (3.5' $\times$ 18'), nearly 50,000 times larger than the projected area of sky covered by the ARCES slit! Therefore any water sublimation from an extended source of icy grains could be missed by the ARCES observations but would be captured by Nan\c{c}ay, explaining the larger production rate measured by Nan\c{c}ay.  The factor of five discrepancy is similar to the discrepancy observed for C/2009 P1 (Garradd) between narrow slit and wide field observations at similar heliocentric distance~\citep{Combi2013,McKay2015}.  The~\cite{Opitom2019b} upper limit on the OH production of 2.0 $\times$ 10$^{27}$ mol s$^{-1}$ is also inconsistent with the reported tentative Nan\c{c}ay detection.  Optitom et al. used a narrow slit spectrometer (2\arcsec $\times$ 8\arcsec) with a field of view much smaller than the Nan\c{c}ay observations, and so their observation is also consistent with an extended source of water production outside the slit of~\cite{Opitom2019b} but inside the field of view of Nan\c{c}ay.  If 2I/Borisov is indeed hyperactive as suggested earlier, an extended source of water production would be expected.  Adopting the Crovisier et al. water production rate would increase our active fractions by a factor of five, making it quite likely that Borisov would have to be hyperactive to explain their observations.  However, a full comparison of the different constraints on water production will have to await publication of the Nan\c{c}ay results.  Additional measurements/constraints on water production are also needed to confirm or refute the hypothesis of an extended source of water production and hyperactivity.

\section{Conclusions}
\indent We present spectra showing detection of [\ion{O}{1}]6300~\AA~emission in interstellar comet 2I/Borisov.  This provides the first measurement of the H$_2$O production rate in this very intriguing object.  We determined an H$_2$O production rate of (6.3$\pm$1.5) $\times$ 10$^{26}$ mol s$^{-1}$, which when compared to CN measurements suggest Borisov is either enhanced or typical in CN compared to the average value for Solar System comets, though the enhanced numbers are still within the range of observed values.  C$_2$ is depleted compared to H$_2$O.  The dust-to-gas ratio based on Af$\rho$ and dust mass estimates are consistent with Solar System comets.  Using a simple sublimation model, we find an H$_2$O active area of 1.7 km$^2$, which for current constraints on the nucleus size could imply active fractions from as low as 1\% to $>$ 100\% (implying a hyperactive nucleus), though these active fractions are highly model dependent.  More measurements are needed as Borisov approaches perihelion to fully understand its composition and activity. 

\acknowledgements
We thank the anonymous reviewer whose comments improved the quality of this manuscript.  Based on observations obtained with the Apache Point Observatory 3.5-meter telescope, which is owned and operated by the Astrophysical Research Consortium. We acknowledge funding from the NASA Solar System Observations Program through grant 18-SSO18-2-0040. 

\facilities{APO-ARCES}

\bibliography{references.bib}
\bibliographystyle{plainnat}

\end{document}